\documentclass[12pt]{article}
\usepackage[latin1]{inputenc}
\usepackage{slashed}
\usepackage{amsmath}
\usepackage{textcomp}
\usepackage{amssymb}
\usepackage{amsfonts}
\usepackage{indentfirst}
\usepackage{color}
\usepackage[top=2cm,bottom=2cm,left=3cm,right=2cm]{geometry}
\newcommand{\nin}{\noindent}
\newcommand{\be}{\begin{equation}}
\newcommand{\ee}{\end{equation}}
\newcommand{\bea}{\begin{eqnarray}}
\newcommand{\eea}{\end{eqnarray}}

\newcommand{\nn}{\nonumber\\}

\usepackage{graphicx}

\begin{document}

\nin KCL-PH-TH/2016-10

\vspace{1cm}

\begin{center} 
 
 {\Large{\bf Convexity at finite temperature\\ and non-extensive thermodynamics}}

\vspace{0.5cm}

{\bf J. Alexandre} 

\vspace{0.2cm}

King's College London, Department of Physics, WC2R 2LS, UK\\
{\small jean.alexandre@kcl.ac.uk}

\vspace{1cm}

{\bf Abstract}

\end{center}

\vspace{0.5cm}
 
Assuming that tunnel effect between two degenerate bare minima occurs, in a scalar field theory at finite volume, this article studies the consequences   
for the effective potential, to all loop orders. Convexity is achieved only if the two bare minima are taken into account in the path integral, and
a new derivation of the effective potential is given, in the large volume limit. The effective potential has then
has a universal form, it is suppressed by the space time volume, and does not feature spontaneous symmetry breaking as long as the volume is finite. 
The finite temperature analysis leads to surprising thermal properties, following from the non-extensive expression for the free energy.
Although the physical relevance of these results is not clear, the potential application to ultra-light scalar particles is discussed.

\section{Introduction}

For a scalar theory with several degenerate vacuua, it is usually assumed that spontaneous symmetry breaking (SSB) occurs and that one specific 
vacuum is chosen. This is actually true for infinite volume, where the tunnel effect between different vacuua is completely suppressed. 
But for finite volume, even the slightest tunneling possibility between different degenerate vacuua should allow these to play
an equivalent role at equilibrium, for the true vacuum of the dressed theory to be a superposition of the bare vacuua.
It has been known for a long time that the effective potential is then convex \cite{Symanzik}, which is a consequence of its definition 
in terms of a Legendre transform \cite{Haymaker}.
It has been shown, from the early days of effective potential methods \cite{Jackiw}, that convexity cannot be achieved when 
quantisation is based on one vacuum only, if the bare potential has several degenerate vacuua  \cite{CJP}.
The effective potential actually becomes convex as a consequence of the competition of the different non-trivial saddle points \cite{Fujimoto}, and
is thus a non-perturbative effect. 

Gauge fixing could also impose a specific vacuum for the scalar field, therefore avoiding convexity of the effective potential.
Nevertheless, a construction of a convex effective Higgs potential is given in \cite{Hindmarsh}, where gauge fixing picks two points on the manifold of vacuua
of the bare potential. It is shown that a convex effective potential can then be obtained from a linear interpolation between these two vacuua, to any loop order.
The explicit form of the effective potential is not given though.

An explicit construction of the convex saddle point effective potential (ignoring loop corrections) is given for the first time in \cite{AT}, where the effective potential is 
derived as an expansion in the classical field, up to the fourth order, for a finite spacetime volume $V^{(4)}$.
In this work, the degenerate vacuua of an $O(N)$-symmetric scalar theory, with $|\vec\phi_{vac}|=v$,
all contribute to the saddle point approximation for the partition function.
The path integral quantisation is then followed step by step, where all the quantities are expanded in either 
the source or the classical field.
The resulting effective potential is a convex polynomial, which is suppressed by $V^{(4)}$, 
as a consequence of an interpolation between the different bare vacuua. Therefore it becomes flat in the 
limit of infinite volume, and SSB is reached only in this limit, 
where the true vacuum is an arbitrary point of a flat $N$-ball with radius $v$. 

Although not studied in \cite{AT}, Goldstone modes then arise, which should stay massless to all orders in perturbation theory. This is shown in \cite{Teresi}, using 
an improved Conwall-Jackiw-Tomboulis effective action \cite{CJT}.

The present article (restricted to a single real scalar field) shows an alternative construction, which is not based on an 
expansion in fields, 
but in $(v^4V^{(4)})^{-1}$ instead. This approach leads to an effective potential which is valid to all orders in the classical field,
and whose Taylor expansion to the fourth order is consistent with \cite{AT}. 
The true vacuum of the (dressed) theory is located at $\phi=0$, and SSB does not occur as long as $V^{(4)}$ is finite.
This result is first obtained in the saddle point approximation for the partition function, at zero temperature, and we show that one-loop corrections do 
not change the functional form of the effective potential, but only redefine the mass scale $v$. We show then that these results hold at finite temperature too, as
long as one is below a critical temperature. 

Concerning the true vacuum of a theory, we note that a systematic construction of effective theories is done in \cite{Skliros}, where tadpoles are removed consistently with 
the true vacuum. In the situation where the bare vacuua are not degenerate, one can also consider the famous problem of false vacuum decay \cite{CaCo},
for which radiative corrections are considered in \cite{Millington}, and gauge invariance is shown in \cite{Tamarit}. But these studies assume a time
dependence of the ground state of the theory, whereas we consider here the equilibrium situation in the case of a symmetric potential. 

This article is structured as follows. The explicit construction of the effective potential at zero temperature, within   
the saddle point approximation, is done in section 2. The one-loop corrections are calculated in section 3, and the results are expected to be identical at higher order loops, up to
a redefinition of the mass scale $v$ of the theory. The extension to finite temperature is done in section 4, where the effective potential is suppressed by the
three-dimensional space volume, as long as one stays below the usual critical temperature. This suppression holds in an interval defined by the temperature-dependent renormalised 
mass scale corresponding to the vev $v$. 
An important consequence is that the free energy is not extensive, as well as the entropy, and the latter happens to be a constant which can be simply interpreted.
The conclusion of the article discusses potential physical applications of this volume-suppressed convex effective potential, which could be relevant in a cosmological context.
The article ends with four Appendices:\\
{\it(i)} Appendix A gives a general argument for convexity of the effective potential;\\
{\it(ii)} Appendix B shows that the effective potential obtained from the Legendre transform, as defined in this article, and the Wilsonian effective potential are identical
in the limit of infinite volume;\\
{\it(iii)} Appendix C shows that bounce saddle points between the two bare vacuua do not play a role for the effective potential, if one starts with a symmetric bare potential;\\
{\it(iv)} Appendix D treats the example of a cosine bare potential, for which it is shown that the saddle point effective potential is completely flat. This result is expected from 
the assumptions of convexity, periodicity and differentiability.

\section{Saddle point approximation}

We start from the bare potential
\be\label{sympot}
U_{bare}(\phi)=\frac{\lambda}{24}(\phi^2-v^2)^2~,
\ee
and we are interested in the effective potential only, such that the source $j$ appearing in the partition function is chosen as a constant. We note, in Euclidean metric,
\be
Z[j]=\int{\cal D}[\phi]\exp\left(-S[\phi]-\int j\phi\right)\equiv e^{-W[j]}~,
\ee
and the partition function $Z[j]$ will be approximated by the sum over the dominant saddle points $\phi_n$.
We will consider the uniform configurations only, since we show in appendix B that the ``bounce'' solutions, originally considered in \cite{CaCo} for the calculation of
tunneling rates, are negligible. \\
The uniform saddle points of the partition function are solutions of the equation
\be\label{equamot}
U_{bare}'(\phi_n)+j=0~,
\ee
and the number of these solutions depends on the source $j$ \cite{AT}:
\begin{itemize}
 
\item if $|j|>j_c$, where $j_c=\lambda v^3/(9\sqrt3)$, then eq.(\ref{equamot}) has one real solution only, that we note $\phi_0(j)$;

\item if $|j|<j_c$, then eq.(\ref{equamot}) has the three solutions, one of which is a maximum and the two others are the local minima relevant for the partition function
\bea\label{phi12}
\phi_1&=&\frac{2v}{\sqrt3}\cos[\pi/3-(1/3)\arccos(j/j_c)]\\
\phi_2&=&-\frac{2v}{\sqrt3}\cos[(1/3)\arccos(j/j_c)]~.\nonumber
\eea

\end{itemize}
In what follows we are looking at these two situations separately.

\subsection{large source}

For the situation where $|j|>j_c$, the saddle point partition function is
\be
Z^{(0)}[j]=\exp\left(-V^{(4)}U_{bare}(\phi_0)-V^{(4)}j\phi_0(j)\right)~.
\ee
Since the source $j$ is uniform, functional derivatives with respect to $j$ are replaced by partial derivatives with respect to $V^{(4)}j$
\be
\frac{\delta }{\delta j}\to\frac{1}{V^{(4)}}\frac{\partial }{\partial j}~,
\ee 
where $V^{(4)}$ is the spacetime volume, such that the classical field is 
\be
\phi_c\equiv\frac{1}{V^{(4)}}\frac{\partial W^{(0)}}{\partial j}=(U'_{bare}(\phi_0)+j)\frac{\partial\phi_0}{\partial j}+\phi_0(j)=\phi_0(j)~.
\ee
The saddle point effective action is then
\be
\Gamma^{(0)}[\phi_c]=W^{(0)}[j]-V^{(4)}j\phi_c=V^{(4)}U_{bare}(\phi_c)~,
\ee
leading to the saddle point effective potential 
\be
U^{(0)}_{eff}(\phi_c)=\frac{\Gamma^{(0)}[\phi_c]}{V^{(4)}}=U_{bare}(\phi_c)~.
\ee
This expected result corresponds to the situation where $\phi_c$ is outside the minima $v_1=v$ and $v_2=-v$: the saddle point approximation does not modify the bare potential.

\subsection{Small source}

For uniform fields, we have seen that the relevant source is actually $k\equiv V^{(4)}j$, which will allow to make an expansion in  
the small parameter $\epsilon\equiv (v^4V^{(4)})^{-1}$. \\
For a small source $|j|<j_{crit}$, the saddle point partition function is given by  
\be\label{Z0small}
Z^{(0)}[k]=\exp\left(-V^{(4)}U_{bare}(\phi_1)-V^{(4)}j\phi_1\right)+\exp\left(-V^{(4)}U_{bare}(\phi_2)-V^{(4)}j\phi_2\right)~,
\ee
where $\phi_{1,2}$ are given by eqs.(\ref{phi12}).
One then expands the arguments of the exponentials in $\epsilon<<1$
\bea\label{initialexpansion}
V^{(4)}U_{bare}(\phi_n)+V^{(4)}j\phi_n&=&V^{(4)}j\left(U_{bare}'(v_n)\left.\frac{\partial\phi_n}{\partial j}\right|_0+v_n\right)+V^{(4)}{\cal O}(j^2)\nn
&=&kv_n+k^2{\cal O}(\epsilon)~,
\eea
where $n=1,2$, and the partition function (\ref{Z0small}) is then
\be\label{Z0resmall}
Z^{(0)}[k]\simeq2\cosh(kv)+{\cal O}(\epsilon)=e^{-W^{(0)}[k]}~.
\ee
Note that such an expansion in $\epsilon$ is not valid when the partition function is dominated by one minimum only: in the situation where $k>0$ for example, 
the term of order $\epsilon$ can actually be dominant $e^{-kv}<<{\cal O}(\epsilon)$, and it is only the sum $e^{-kv}+e^{kv}$ which is always 
large compared to ${\cal O}(\epsilon)$, whatever the sign of $k$ is.

Neglecting terms of order $\epsilon$, the classical field obtained from eq.(\ref{Z0resmall}) is then 
\be\label{phic}
\phi_c=\frac{\partial W^{(0)}}{\partial k}=-v\tanh(kv)~,
\ee
and the latter relation can be inverted to obtain
\be\label{kphic}
k=\frac{1}{2v}\ln\left(\frac{v-\phi_c}{v+\phi_c}\right)~.
\ee
The saddle point effective action is then
\be
\Gamma^{(0)}[\phi_c]=-\ln\left(2\cosh(kv)\right)-k\phi_c~,
\ee
and the relation (\ref{kphic}) gives
\be
\Gamma^{(0)}[\phi_c]=\frac{1}{2}\left(1+\frac{\phi_c}{v}\right)\ln\left(1+\frac{\phi_c}{v}\right)
+\frac{1}{2}\left(1-\frac{\phi_c}{v}\right)\ln\left(1-\frac{\phi_c}{v}\right)-\ln2~.
\ee
The saddle point effective potential is finally obtained after dividing by the spacetime volume
\be\label{semiclass}
U_{eff}^{(0)}(\phi_c)=-\frac{\ln2}{V^{(4)}}+\frac{1}{2V^{(4)}}\left(1+\frac{\phi_c}{v}\right)\ln\left(1+\frac{\phi_c}{v}\right)
+\frac{1}{2V^{(4)}}\left(1-\frac{\phi_c}{v}\right)\ln\left(1-\frac{\phi_c}{v}\right)~.
\ee
The latter potential is convex, and matches the bare potential at $\phi_c=\pm v$, leading to an overall continuous saddle point effective potential. 
For small values of the field $|\phi_c|<<v$, a Taylor expansion gives
\be\label{UeffAT}
U_{eff}^{(0)}(\phi_c)=-\frac{\ln2}{V^{(4)}}+\frac{1}{2V^{(4)}}\left(\frac{\phi_c}{v}\right)^2+\frac{1}{12V^{(4)}}\left(\frac{\phi_c}{v}\right)^4+\cdots
\ee
which was found in \cite{AT}, apart from the constant shift $-\ln2/V^{(4)}$, arising from an overall factor 1/2 in the partition function defined in \cite{AT}. In flat spacetime, 
this shift is not relevant, but in curved space time, if one imposes a vanishing vacuum 
energy in the bare potential, this shift induces a vacuum energy in the effective theory, for the true vacuum $\phi_c=0$.

One can note that the approach given in \cite{AT} is based on an expansion in the source up to the order $j^4$, and therefore $\phi_c^4$,
but taking into account all the orders in $(V^{(4)})^{-1}$. The approach adopted here, on the other hand, 
is based on an expansion in $(V^{(4)})^{-1}$, which has the advantage to provide the resummation (\ref{semiclass}) to all orders in the classical field $\phi_c$.

The effective potential (\ref{semiclass}) is universal in the sense that it depends on the bare vev $v$ only, 
and not on the explicit form of the bare potential. The coupling constant $\lambda$ of the original model (\ref{sympot}) does not appear in the final expression 
(\ref{semiclass}), because it cancels out in the limit of large volume, for which the large parameter is $\lambda V^{(4)}v^4/24\propto\lambda/\epsilon$, 
as shown in \cite{AT}. Therefore, as expected, the present construction is not valid in the limit $\lambda\to0$.

Finally, we also note that, although continuous at $\phi_c=\pm v$, the effective potential (\ref{semiclass}) is not analytical at these points. This
is because the limits $\phi_c\to\pm v$ correspond to the transition between small and large source, where the saddle point approximation for the partition function is not sufficient.
As explained in \cite{Tamarit}, in this situation there is an overlap of the wave functions corresponding to the two ground states dominating the partition function, 
which is then not well approximated by the sum of two independent terms.

\subsection{Maxwell construction and Wilsonian approach}

For finite spacetime volume, the true minimum of the system is at $\phi_c=0$,
as a result of a tunnel effect between the two bare minima. In the limit of infinite spacetime volume $V^{(4)}\to\infty$ though, the effective potential (\ref{semiclass}) 
is exactly flat between the bare vacuua, the true vacuum is not unique anymore, and SSB occurs. 
The vacuum of the system consists then in a superposition of 
the two bare vacuua, with a weighted average $\phi_c$ which can be in an arbitrary position between the bare vacuua. 

This situation is similar to the so-called Maxwell construction in the study of the Van der Waals equation of state, where naive isothermal curves in the plane $(P,V)$ present a region
of negative compressibility, which is not physical. The origin of the problem is that the Van der Waals equation of state describes only one phase of the fluid, and 
the solution to the problem is to split the fluid in two phases, liquid and vapour. There is then only one independent
state variable, and since the temperature $T$ is already fixed, the pressure is also fixed at the saturated vapour pressure $P_S(T)$: 
the true isotherm trajectory features a plateau,
avoiding the negative compressibility. 

This plateau is representative of the coexistence of two phases, and the classical analogy in the present case is the 
following \cite{Miransky}: the flat potential in the 
limit $V^{(4)}\to\infty$ corresponds to the coexistence of bubbles of different vacuua $\pm v$, with arbitrary sizes.

\vspace{0.5cm}

The effective potential defined in the present article corresponds to the one-particle irreducible - 1PI - effective potential. 
It is identical to the Wilsonian effective potential in the limit of infinite volume only (see Appendix B),
such that there are similarities and differences with the latter, few of which are discussed here.

Convexity in the Wilsonian approach was originally obtained in \cite{Wetterich} through the contribution of a non-trivial
saddle point to the evolution equation, in the framework of the average effective action. 
Similarly, using a sharp cut off though, the contribution of a non-trivial saddle point in every infinitesimal blocking step lead to 
the explicit construction of the flat Wilsonian effective potential \cite{ABP}. The Wilsonian approach consists in averaging over 
different microscopic configurations,
with momentum typically larger than some scale $k$, 
and gradually build the effective potential in the infrared limit $k\to0$ (IR). 
In the present work, averaging over different microscopic configurations is the essential point at the origin of convexity, 
as explained previously, which reflects in a way the Wilsonian approach.
But the analogy must be taken carefully, for the following reasons:\\
{\it(i)} The Wilsonian running potential is convex in the limit $k\to0$ only;\\
{\it(ii)} The convex Wilsonian effective potential (for $k=0$) is flat for any space time volume, 
whereas the 1PI effective potential considered here, although convex for any volume, is flat for $V^{(4)}\to\infty$ only;\\
{\it(iii)} The semi-classical construction of the partition function (\ref{Z0small}) averages over two classical vacuua, 
without the contribution of quantum fluctuations. On the other hand, the Wilsonian approach automatically involves 
quantum fluctuations above the two vacuua. In the present construction, 
convexity is already obtained at the semi-classical level, and quantum fluctuations only modify the mass parameter of the model 
(see next section).

\section{One-loop corrections (zero temperature)}

We show here that the overall picture is not modified by one-loop corrections, apart from corrections to the mass scale $v$. Higher order loop 
corrections would have the same effect, and the existence of the volume-suppressed part of the effective potential is expected to be an exact result.

\subsection{Large source}

In the situation of large source described in subsection 2.1 for the saddle point approximation, the one-loop corrections are obtained by considering 
quadratic fluctuations around the minimum $\phi_0(j)$
\be\label{Z1}
Z^{(1)}_{out}[j]=e^{-V^{(4)}U_{bare}(\phi_0)-V^{(4)}j\phi_0}\int{\cal D}[\xi]\exp\left(-\frac{1}{2}\int\int\xi\xi\delta^2S|_{\phi_0}\right)~,
\ee
and the usual steps of path integral quantisation lead to the known expression
\bea\label{out}
U^{(1)}_{out}(\phi_c)&=&U_{bare}(\phi_c)+\frac{1}{2V^{(4)}}\mbox{Tr}\left\{\ln(\delta^2S|_{\phi_c})\right\}\\
&=&\frac{\lambda}{24}(\phi_c^2-v^2)^2+\frac{1}{2}\int\frac{d^4p}{(2\pi)^4}\ln\left(\frac{p^2-\lambda v^2/6+\lambda\phi_c^2/2}{p^2+M^2}\right)~,\nonumber
\eea
where $M$ is an arbitrary mass scale defining the zero of the potential. But this expression is consistent as long as 
$\phi_c^2>v^2/3$ only, which corresponds to the inflexion points of the bare potential. It becomes singular when $\phi_c^2\to v^2/3$, and 
contains an imaginary part for
$\phi_c^2<v^2/3$, which is the sign that some wrong step was taken.

The problem with the expression (\ref{out}) when $\phi_c^2\leq v^2/3$ is the cancellation of restoration force for 
quantum fluctuations $\xi$ with momentum $p_\mu$ satisfying $p^2-\lambda v^2/6+\lambda\phi_0^2/2<0$ in the path integral (\ref{Z1}), 
such that this partition function is not correct anymore. As shown in the next subsection, the problem is cured by taking into account fluctuations around both 
saddle points of the partition function.

But the expression (\ref{out}) can be used to calculate the one-loop correction $v^{(1)}$ 
to the mass scale $v$, since it is perturbatively away from the bare vev and thus must satisfy $v^{(1)}>v/\sqrt3$. 
We have then
\be
0=\left.\frac{dU^{(1)}_{out}}{d\phi_c}\right|_{v^{(1)}}=\frac{\lambda}{6}v^{(1)}[(v^{(1)})^2-v^2]
+\frac{\lambda}{2}v^{(1)}\int\frac{d^4p}{(2\pi)^4}\frac{1}{p^2}+{\cal O}(\lambda^2)~,
\ee
such that
\be\label{v1}
(v^{(1)})^2\simeq v^2\left(1-\frac{3\Lambda^2}{16\pi^2v^2}\right)~,
\ee
where $\Lambda$ is an ultraviolet cut off. Although not obviously visible from the latter expression, quantum corrections are indeed perturbative:
if the bare potential is written
\be
\frac{\lambda}{24}(\phi^2-v^2)^2=\frac{\lambda}{24}v^4-\frac{\mu^2}{2}\phi^2+\frac{\lambda}{24}\phi^4~,
\ee
then $v^2=6\mu^2/\lambda$, such that the correction in $\Lambda^2/v^2$ is actually proportional to $\lambda\Lambda^2/\mu^2$.  
As we will see below, the expression (\ref{v1}) is also obtained from the one-loop effective potential for small source.

\subsection{Small source}

We show here that one-loop corrections do not change the functional form of the effective potential (\ref{semiclass}), but only redefine the value of the mass scales $v_n$.
An important message is that there is no imaginary part generated by quantum corrections, because of the interpolation between the two vacuua.

One-loop quantum corrections to the effective potential (\ref{semiclass}) are obtained after taking into account quadratic fluctuations around each saddle points
\bea\label{Zdebase}
Z^{(1)}_{in}[j]&=&\sum_n e^{-V^{(4)}U_{bare}(\phi_n)-V^{(4)}j\phi_n}\int{\cal D}[\xi]\exp\left(-\frac{1}{2}\int\int\xi\xi\delta^2S|_{\phi_n}\right)\nn
&=&\sum_n\exp\left\{-V^{(4)}U_{bare}(\phi_n)-V^{(4)}j\phi_n-\frac{V^{(4)}}{2}\int_p\ln\left(\frac{p^2+U_{bare}''(\phi_n)}{p^2+U_{bare}''(v_n)}\right)\right\}~.
\eea
As shown in the previous section, only the linear term in $j$ is relevant for the first order in $(V^{(4)})^{-1}$. We have 
\bea
&&U_{bare}(\phi_n)+j\phi_n+\frac{1}{2}\int_p\ln\left(\frac{p^2+U_{bare}''(\phi_n)}{p^2+U_{bare}''(v_n)}\right)\nn
&=&j\left(v_n+\frac{1}{2}U_{bare}'''(v_n)\left.\frac{\partial\phi_n}{\partial j}\right|_0\int_p\frac{1}{p^2+U_{bare}''(v_n)}\right)+{\cal O}(j^2)~,
\eea
which, compared to eq.(\ref{initialexpansion}), shows that the only change to the effective potential (\ref{semiclass}) is a redefinition of the mass scales $v_n$
\be\label{vn1}
v_n\to v_n^{(1)}=v_n+\frac{1}{2}U_{bare}'''(v_n)\left.\frac{\partial\phi_n}{\partial j}\right|_0\int_p\frac{1}{p^2+U_{bare}''(v_n)}~.
\ee
One can conjecture that any higher order loop correction would have the same effect, with a different coefficient for the first order of the expansion in the source $j$.\\
From the minima (\ref{phi12}) we have  
\be
\left.\frac{\partial\phi_1}{\partial j}\right|_0=\left.\frac{\partial\phi_2}{\partial j}\right|_0=-\frac{3}{\lambda v^2}~,
\ee
and eq.(\ref{vn1}) leads then to a result consistent with eq.(\ref{v1})
\be\label{v1again}
v_2^{(1)}=-v_1^{(1)}=v-\frac{3}{2v}\int\frac{d^4p}{(2\pi)^4}\frac{1}{p^2+\lambda v^2/3}\simeq v\left(1-\frac{3\Lambda^2}{32\pi^2 v^2}\right)~.
\ee
The overall one-loop effective potential is therefore continuous and convex:
\begin{itemize}
\item For small fields $\phi_c^2<(v^{(1)})^2$, the one-loop effective potential $U^{(1)}_{in}(\phi_c)$ is given by eq.(\ref{semiclass}), with the change $v\to v^{(1)}$;
\item For large fields $\phi_c^2>(v^{(1)})^2$, the one-loop effective potential $U^{(1)}_{out}(\phi_c)$ is given by eq.(\ref{out}), 
with $M^2$ chosen in such a way that $U^{(1)}_{out}(v^{(1)})=0$, i.e.
\be
M^2\simeq\frac{\lambda v^2}{3}\left(1-\frac{9\Lambda^2}{64\pi^2 v^2}\right)~.
\ee
\end{itemize}
A convex one-loop effective potential therefore arises naturally from consistently taking into account both minima of the bare potential, 
from the very beginning of
path integral quantisation. 

We finally note an apparent mismatch of the field-range for which to choose either the potential $U^{(1)}_{in}(\phi_c)$ or $U^{(1)}_{out}(\phi_c)$. Indeed,
the expression $U^{(1)}_{out}(\phi_c)$ is still mathematically consistent for $v^2/3<\phi_c^2<v^2$, although in this range the true effective potential is 
$U^{(1)}_{in}(\phi_c)$ (we neglect in this discussion the corrections to the bare vacuua). 
The reason is that an effective potential which is both convex and differentiable can be obtained only if the volume-suppressed part $U^{(1)}_{in}(\phi_c)$
holds in the whole range between the two bare vacuua. This is confirmed by the Wilsonian approach, which takes
into account non-trivial saddle points in each infinitesimal blocking step \cite{Wetterich,ABP}.

\section{Finite temperature analysis}

At finite temperature $T=\beta^{-1}$, the one-loop analysis is similar, besides the fact that the integration over frequencies $p_0$ 
is replaced the summation over discrete Matsubara modes $\omega_n=2\pi n\beta^{-1}$
\be
\int\frac{d^4p}{(2\pi)^4} f(p_0,\vec k)~\rightarrow~\beta^{-1}\int\frac{d\vec k}{(2\pi)^3}\sum_{n=-\infty}^\infty f(\omega_n,\vec k)~,
\ee
where $p_\mu=(p_0,\vec k)$. 
The saddle point approximation involves only the classical theory, and is therefore identical to the one described in section 2, with the 
replacement $V^{(4)}\to \beta V^{(3)}$, where $V^{(3)}$ is the space volume. In what follows we therefore directly go to one-loop 
corrections.

\subsection{Large source}

Following the steps described in section 3, the one-loop effective potential is, in the situation of large source 
\be\label{UTout}
U^{(1)}_{out}(\phi_c)
=\frac{\lambda}{24}(\phi_c^2-v^2)^2+\frac{\beta^{-1}}{2}\int\frac{d\vec k}{(2\pi)^3}\sum_{n=-\infty}^\infty
\ln\left(\frac{\omega_n^2+k^2-\lambda v^2/6+\lambda\phi_c^2/2}{\omega_n^2+k^2+M^2}\right)~,
\ee
and the one-loop correction $v^{(1)}$ to the mass scale $v$ is given by, in the large-temperature regime $\lambda v^2\beta^2<<1$,
\be
0=\left.\frac{dU^{(1)}_{out}}{d\phi_c}\right|_{v^{(1)}}=\frac{\lambda}{6}v^{(1)}[(v^{(1)})^2-v^2]
+\frac{\lambda}{2}v^{(1)}\beta^{-1}\int\frac{d\vec k}{(2\pi)^3}\sum_{n=-\infty}^\infty\frac{1}{\omega_n^2+k^2}
+{\cal O}(\lambda^2 v^2\beta^2)~.
\ee
The summation over Matsubara modes is done using the known identity
\be
\sum_{n=-\infty}^\infty\frac{1}{n^2+a^2}=\frac{\pi}{a}\coth(\pi a)=\frac{\pi}{a}\left(1+\frac{2}{e^{2\pi a}-1}\right)~,
\ee
and leads to, up to corrections of order $\lambda v^2\beta^2$,
\be
(v^{(1)})^2=v^2-\frac{3}{4\pi^2}\left(\int_0^\Lambda kdk+2\int_0^\infty\frac{kdk}{e^{\beta k}-1}\right)~.
\ee
The first integral corresponds to the zero-temperature correction, whereas the second integral corresponds to the 
finite-temperature contribution, which is not divergent:
\be\label{v1T}
(v^{(1)})^2=v^2-\frac{3\Lambda^2}{8\pi^2}-\frac{1}{4\beta^2}~.
\ee
We note that the zero-temperature correction is twice the one obtained in eq.(\ref{v1}), which is due to a 
different regularisation of the loop integral. 
Indeed, at finite temperature, there is no restriction on the amplitude of the Matsubara modes, whereas at 
zero temperature the integration over frequencies is 
restricted by the cut off.\\
One then defines the renormalised zero-temperature mass scale $v_0$ by
\be
v_0^2\equiv v^2\left(1-\frac{3\Lambda^2}{8\pi^2v^2}\right)~,
\ee
and the corresponding renormalised finite-temperature mass scale is thus given by,
\be\label{vT2}
v_T^2\equiv(v^{(1)})^2=v_0^2-\frac{T^2}{4}~.
\ee
From the previous result, one can define the critical temperature 
\be\label{Tc}
T_c\equiv 2v_0~,
\ee
at which $v_T\to0$, and the transition to larger temperatures is discussed further down.\\
We finally note that the temperature-dependent part of the effective potential (\ref{UTout}) is, for high temperatures \cite{finiteT}, 
\be\label{expT}
U^{(1)}_{out~T}=-\frac{\pi^2T^4}{90}+\lambda\frac{T^2}{48}(\phi^2-3v^2)+\cdots~
\ee
where dots denote higher order terms in $\lambda^{1/2}\beta v$.

\subsection{Small source}

The effective potential (\ref{UTout}), derived for large source, is valid for $|\phi_c|\ge v_T$. 
As discussed in section 3, for $|\phi_c|<v_T$, quantum corrections consist in replacing the mass scale $v$ by its renormalised value, which is $v_T$ here, in the 
volume-suppressed effective potential (\ref{semiclass}). 
Replacing also $V^{(4)}$ by $\beta V^{(3)}$, the effective potential for $|\phi_c|<v_T$ is then:
\be\label{UTin}
U_{in}^{(1)}(\phi_c)=-\frac{T\ln2}{V^{(3)}}+\frac{T}{2V^{(3)}}\left(1+\frac{\phi_c}{v_T}\right)\ln\left(1+\frac{\phi_c}{v_T}\right)
+\frac{T}{2V^{(3)}}\left(1-\frac{\phi_c}{v_T}\right)\ln\left(1-\frac{\phi_c}{v_T}\right)~.
\ee
One can note that the expression for the renormalised temperature-dependent mass scale $v_T$ can also be obtained
for small source. Indeed, the finite-temperature one-loop correction (\ref{vn1}) becomes
\bea
v^{(1)}&=&v+\frac{1}{2}U_{bare}'''(v)\left.\frac{\partial\phi_1}{\partial j}\right|_0~\beta^{-1}\int\frac{d\vec k}{(2\pi)^3}
\sum_{n=-\infty}^\infty\frac{1}{\omega_n^2+k^2+U_{bare}''(v)}\\
&=&v-\frac{3\beta^{-1}}{4\pi^2v}\int_0^\Lambda k^2dk\sum_{n=-\infty}^\infty\frac{1}{\omega_n^2+k^2}+{\cal O}(\lambda v^2\beta^2)~.
\eea
and the summation over Matsubara modes leads to 
\be
v^{(1)}=v-\frac{3\Lambda^2}{16\pi^2v}-\frac{1}{8\beta^2v}~,
\ee
which is consistent with the result (\ref{v1T}) (one must keep in mind that, in the perturbative context, $v$ is large
compared to quantum corrections).

\subsection{Zeroth order phase transition}

The effective potential 
features the volume-suppressed part (\ref{UTin}) in the interval $[-v_T,v_T]$ as long as $T<T_C$. 
As the temperature increases and reaches the critical temperature $T\to T_c$, this interval shrinks and vanishes:
the effective potential (\ref{UTout})
becomes then valid for all the values of the classical field. In both temperature regimes though, the ground state is $\phi_c=0$, provided the volume $V^{(3)}$ is finite.
The limit $T\to T_c$ thus does not corresponds to a spontaneous symmetry breaking for the vacuum, but rather to a different scaling with the volume $V^{(3)}$.

More specifically, let us study the effective mass, the free energy and the entropy, defined in the ground state $\phi_c=0$ (or vanishing source $j=0$):
\bea\label{defmS}
m_T^2&=&\left.\frac{\partial^2U_{eff}}{\partial\phi_c^2}\right|_0\\
F&=&-T\ln Z[0]=V^{(3)}U_{eff}(0)\nn
S&=&-\left(\frac{\partial F}{\partial T}\right)_{V^{(3)}}=-V^{(3)}\frac{\partial U_{eff}(0)}{\partial T}~,\nonumber
\eea
where the Boltzmann constant is set to 1. The transition is of zeroth order, since the free energy $F$ is discontinuous:
\\

$\bullet$ \underline{For $T\ge T_c$}, the effective potential is given by the expression (\ref{UTout}), and field fluctuations around 
the vacuum $\phi_c=0$ see the effective mass with the expected form 
\bea
m_>^2&=&\left.\frac{\partial^2U^{(1)}_{out}}{\partial\phi_c^2}\right|_0
\simeq-\frac{\lambda v^2}{6}+\frac{\lambda\beta^{-1}}{2}\int\frac{d\vec k}{(2\pi)^3}
\sum_{n=-\infty}^\infty\frac{1}{\omega_n^2+k^2}\nn
&=&-\frac{\lambda v^2}{6}\left(1-\frac{3\Lambda^2}{8\pi^2 v^2}\right)+\frac{\lambda}{24\beta^2}
=\frac{\lambda}{24}(T^2-T_c^2)~,
\eea
which vanishes in the limit $T\to T_c$.\\
The free energy and the entropy of the ground state $\phi_c=0$ are obtained from the high-temperature expansion (\ref{expT}), whose leading term 
gives the known expressions
\bea
F_>&=&V^{(3)}U^{(1)}_{out~T}(0)=-\frac{\pi^2}{90}V^{(3)}T^4\\
S_>&=&\frac{2\pi^2}{45}V^{(3)}T^3~;\nonumber
\eea
\\
 
$\bullet$ \underline{For $T<T_c$}, the effective potential features the volume-suppressed part (\ref{UTin}) in the interval $[-v_T,v_T]$.
The effective renormalised mass, defined above the true vacuum $\phi_c=0$, is given by
\be
m_<^2=\left.\frac{\partial^2U^{(1)}_{in}}{\partial\phi_c^2}\right|_0
=\frac{T}{v_T^2V^{(3)}}=\frac{4T}{(T_c^2-T^2)V^{(3)}}~,
\ee 
and diverges at the critical temperature, if the volume $V^{(3)}$ is finite. This divergence could be an artifact arising from the weakness of the  
saddle point approximation for $|\phi_c|\to v_T$, and a more detailed study would be necessary to investigate this limit.\\
The free energy and the entropy, obtained from the potential (\ref{UTin}), read
\bea\label{FS<}
F_<&=&V^{(3)}U^{(1)}_{in~T}(0)=-T\ln2\\
S_<&=&\ln2~,\nonumber
\eea
and are not proportional to the volume, as a consequence of the volume-suppressed form of the effective potential (\ref{UTin}). The expression (\ref{FS<}) for $S$
is the Boltzmann entropy for a system with two degenerate microscopic states, which correspond to the two bare vacuua.\\
Due to the expressions (\ref{FS<}) under the critical temperature, both the pressure and the internal energy vanish in the ground state.
Therefore, in the low temperature regime $T<T_c$, the thermodynamical properties of the system are frozen, due to the specific form (\ref{UTin})
of the effective potential, where only the parameter $v_T$ is modified by quantum corrections. Also, the effective potential (\ref{UTin}) has been derived
independently of a large-temperature assumption, such that its features should remain valid for all temperatures below $T_c$.

\section{Conclusion: physical relevance?}

As we have seen, convexity arises from the interplay of the two degenerate minima of the bare potential, when both are taken into account in the definition of the 
partition function. This is the reason why the Coleman-Weinberg potential (\ref{out}) is different, since it is based on the quantisation over one minimum only.
The present article assumes a finite volume and tunnel effect between the two minima of the bare potential:   
quantisation leads then to a convex effective potential, without imaginary part and without SSB.

But the order in which quantisation is done and the volume is taken to infinity is important: if one first assumes an infinite volume, then tunnel effect is 
completely suppressed and quantisation occurs above one minimum only. Convexity is then not a property of the effective theory, 
because the partition function is a partial one: it does not take into account the whole space of field configurations, and the proof shown in Appendix A is not valid.
In this case, the partial partition function takes into account fluctuations above one ground state only, consistently with SSB. 

It is therefore not obvious to see in what physical situation the present construction can be relevant. 
Nevertheless, a potential application for the dynamical generation of ultra-light scalar Dark Matter \cite{Sin} is proposed in \cite{Alexandre}. 
In order to obtain a coherence length of the size of a typical galactic halo, the mass of these particles should typically be of the order $10^{-23}$ eV. It has been 
 shown in \cite{Alexandre} that such a mass is provided by the effective potential (\ref{UeffAT}), where $v$ is the Higgs vev and $V^{(4)}=L^4$, with $L$
 the particle horizon at the time of the Electroweak phase transition. A common mechanism with the Higgs mechanism is therefore proposed, which could explain how such a small 
 mass can arise from a typical Standard Model mass scale. The extension to finite temperature is planned for a future work.

\vspace{1cm}

\nin{\bf Acknowledgements} I would like to thank Peter Millington, Dimitris Sklyros, Carlos Tamarit, Daniele Teresi and Antonis Tsapalis for discussions on the topic of convexity.

\section*{Appendix A: Convexity of the effective potential}

The proof presented here is taken from \cite{Haymaker}, and assumes a summation over all the field configurations for the calculation of the partition function, 
which allows for a tunnel effect between the different vacuua, if there are. 

Starting from the bare action $S[\phi]$, the (Euclidean) partition function is
\be
Z[j]=\int{\cal D}[\phi]\exp\left(-S[\phi]-\int_x j\phi\right)~,
\ee
where $j(x)$ is the source. The connected graph generating functional is then 
\be
W[j]=-\ln(Z[j])~,
\ee
from which the classical field is defined as
\be\label{classicphi}
\phi_c(x)=\frac{\delta W}{\delta j(x)}=\left<\phi(x)\right>~,
\ee
where
\be
\left<\cdots\right>\equiv\frac{1}{Z}\int{\cal D}[\phi](\cdots)\exp\left(-S[\phi]-\int j\phi\right)~,
\ee
The one-particle irreducible (1PI) graph generating functional $\Gamma[\phi_c]$ is defined as the Legendre transform of $W[j]$
\be
\Gamma[\phi_c]=W[j]-\int_x j\phi_c~,
\ee
where $j$ should be seen a functional of $\phi_c$, after inverting the definition (\ref{classicphi}).
From the definition of the Legendre transform $\Gamma$, one can write the equation of motion for the dressed system as
\be
\frac{\delta\Gamma}{\delta\phi_c}=\int_x\frac{\delta W}{\delta j}\frac{\delta j}{\delta \phi_c}
-\int_y\frac{\delta j}{\delta \phi_c}\phi_c-j=-j~,\nonumber
\ee
and a further derivative gives
\be\label{d2Gd2Wbis}
\frac{\delta^2\Gamma}{\delta\phi_c\delta\phi_c}=-\frac{\delta j}{\delta\phi_c}=-\left(\frac{\delta^2 W}{\delta j\delta j}\right)^{-1}~.
\ee
But one also has
\be
\frac{\delta^2W}{\delta j\delta j}=\left<\phi\right>\left<\phi\right>-\left<\phi\phi\right>~.
\ee
which shows that the second functional derivative of $W$ is necessarily negative, as the opposite of a variance squared.
As a consequence, and given the relation (\ref{d2Gd2Wbis}), the second functional derivative of $\Gamma$ is positive: 
the 1PI effective action is a convex functional.
Its derivative-independent part, the 1PI effective potential, is thus a convex function: $U_{eff}''(\phi_c)\ge0$ everywhere.

\section*{Appendix B: Equivalence between effective potentials}

The Wilsonian effective potential is defined as 
\be
\exp\left( iV^{(4)}U_{Wils}(\phi_0)\right)=\int{\cal D}[\phi]\delta\left(\int_x(\phi-\phi_0)\right)\exp\left( iS[\phi]\right)~,
\ee
and the Dirac distribution is then written as the Fourier transform of an exponential
\bea
\exp\left( iV^{(4)}U_{Wils}(\phi_0)\right) &=&\int dj\int{\cal D}[\phi]\exp\left( iS[\phi]+ij\int_x (\phi-\phi_0)\right)\nn
&=&\int dj \exp\left(iW[j]-ijV^{(4)}\phi_0\right)~. 
\eea
The integration over $j$ is evaluated with the saddle point approximation, which is exact in the limit $V^{(4)}\to\infty$:
\be
\lim_{V^{(4)}\to\infty}\exp\left( iV^{(4)}U_{Wils}(\phi_0)\right)=\lim_{V^{(4)}\to\infty}\exp\left(iW[j_0]-ij_0V^{(4)}\phi_0\right)~,
\ee
where $j_0$ satisfies $\delta W/\delta j_0=\phi_0$, such that $\phi_0$ is the classical field corresponding to $j_0$ and
\be
\lim_{V^{(4)}\to\infty}\exp\left( iV^{(4)}U_{Wils}(\phi_0)\right)
=\lim_{V^{(4)}\to\infty}\exp\left( iV^{(4)}U_{1PI}(\phi_0)\right)~.
\ee
This shows the equivalence between $U_{Wils}$ and $U_{1PI}$, in the limit where $V^{(4)}\to\infty$, and
we note that this argument is valid with a Minkowski metric only, since 
it is based on the Fourier transform of the Dirac distribution.

\section*{Appendix C: (no) Contribution of bounce saddle points}

The aim of this appendix is to show that the contribution of non-homogeneous saddle points is negligible compared to the uniform configurations considered 
in this article, from which the convex potential is obtained. In what follows, we use several aspects described in \cite{CaCo} for the calculation 
of tunneling rates.\\

Any kink-like saddle point of the partition function, which does not depend on at least one of the 4 Euclidean spacetime coordinates, gives rise to a
kinetic term which is proportional to the size of spacetime in the corresponding dimension. As a consequence such a configuration is negligible compared to 
the two uniform saddle points, for which the kinetic term vanish. \\
A non-homogeneous saddle point $\phi_b$ which could potentially contribute to the partition function therefore necessarily has a finite 4-dimensional extension, 
it depends on the $O(4)$-invariant argument $r=\sqrt{x^2+y^2+z^2+t^2}$, and reaches the two different minima of the bare potential for $r=0$ and $r\to\infty$. 
Thus the limits $t\to\pm\infty$ correspond to $r\to\infty$ for the field $\phi_b$, which then starts from its initial value $v$ ($t\to-\infty$), 
goes to its intermediate value $\simeq-v$ ($t=r=0$) and ``bounces'' back to its initial value $v$ ($t\to+\infty$).   
In the absence of source, the bounce satisfies the equation of motion
\be\label{equamotnonhomo}
\frac{1}{r^3}\partial_r(r^3\partial_r\phi_b)=\frac{\lambda}{6}\phi_b(\phi_b^2-v^2)~,
\ee
and corresponds to a 4-dimensional bubble of arbitrary radius $R$. In the thin bubble wall limit $Rv\sqrt\lambda>>1$, an approximate family of solution of eq.(\ref{equamotnonhomo}),
parametrised by the radius $R$, is given by
\be\label{bubble}
\phi_b=v\tanh\left(\frac{r-R}{r_0}\right)~,~~~~~~~r_0=\frac{\sqrt{12}}{v\sqrt\lambda}~.
\ee
In this approximation, the bounce action is dominated by the kinetic term and reads
\be
B\equiv S[\phi_b]\simeq\frac{\pi^2\sqrt\lambda}{6}(vR)^3>>1~,
\ee
which thus gives a negligible contribution to the partition function. As 
explained more generally in \cite{CaCo}, the radius of the bounce solutions which minimises the bounce action is proportional to the inverse of the
difference in energy $\Delta U$ corresponding to the two vacuua. The bounce action $B$ is then proportional to $(\Delta U)^{-3}$, and becomes infinite
in the symmetric limit we consider here, therefore not contributing to the saddle point approximation for the partition function.\\
The bounce solution is actually invariant under spacetime translation, and one should sum over the positions of the centre of the bounce.
As shown in \cite{CaCo}, the contribution of $n$ bounces, taking into account the different locations 
of the centres of each bubble, leads to the final contribution to the partition function of the form 
\be
Z_b\simeq\exp(e^{-B})-1~.
\ee
The latter expression does not take into account the case $n=0$, and is therefore valid for one bounce at least. 
As can be seen, when $\Delta U\to0$ and thus $B\to\infty$, then $Z_b\to0$: bounces do not play a role in the situation of a symmetric bare potential.

\section*{Appendix D: Cosine bare potential}

We consider here the bare potential, which is twice differentiable, 
\bea\label{cosine}
U_{bare}(\phi)&=&M^4\left[1+\cos(\phi/f)\right]~~~~~~~~~~~~~~~\mbox{for}~~~|\phi|\leq(2N+1)\pi f\\
U_{bare}(\phi)&=&\frac{M^4[\phi^2-(2N+1)^2\pi^2 f^2]^2}{8(2N+1)^2\pi^2f^4}~~~~~\mbox{for}~~~|\phi|\geq(2N+1)\pi f~,\nonumber
\eea
where $M$ and $f$ are two mass scales. This potential leads to a converging path integral, since it provides 
quartic restoration forces for large fluctuations $ |\phi|>>(2N+1)f$. 
For finite $N$ we follow the same steps as those shown in section 2.2 to find the classical field in terms of the source $k=V^{(4)}j$. 
In order to recover the full cosine potential, we will then take the limit $N\to\infty$, where it is shown that 
the effective potential becomes flat.

The minima of the bare potential (\ref{cosine}) are given by $\phi_n=(2n+1)\pi f$, for the integers $|n|\leq N$. The saddle point partition function is then, 
up to terms proportional to $(V^{(4)})^{-1}$,
\bea
Z_N^{(0)}&=&\sqrt{q}\sum_{n=0}^N q^n+\sqrt{q^{-1}}\sum_{n=0}^N (q^{-1})^n~~,~~~~~~~~~~~~q\equiv\exp(-2\pi fk)~,\\
&=&\frac{1-q^{N+1}}{q^{-1/2}-q^{1/2}}+\frac{1-q^{-N-1}}{q^{1/2}-q^{-1/2}}\nn
&=&\frac{\sinh[2(N+1)x]}{\sinh(x)}~~,~~~~~~~~~~~~~~~~~~~~~~~~~~~~~x\equiv\pi kf~.\nonumber
\eea
The classical field is then given by
\be
\frac{\phi_c}{\pi f}=\frac{-1}{Z_N^{(0)}}\frac{dZ^{(0)}_N}{dx}=-2(N+1)\coth[2(N+1)x]+\coth(x)~,
\ee
which, in the large $N$ limit reads  
\be
\frac{\phi_c}{\pi f}+2s(N+1)\simeq\coth(\pi kf)~~,~~~~s\equiv~\mbox{sign}(k)~~,~~~N>>1~.
\ee
Inverting this relation gives
\be
2\pi fk=\ln\left(\frac{[2s(N+1)+1]\pi f+\phi_c}{[2s(N+1)-1]\pi f+\phi_c}\right)~,
\ee
which, in the limit where $N\to\infty$ implies $k=0$. Although the source is in principle a free parameter, we can see that the limit $N\to\infty$ 
implies a strong constraint on the system, which shows that the one-to-one mapping between
the source and the classical field is lost. From the generic relation
\be
\frac{\partial\Gamma}{\partial\phi_c}=-k~,
\ee
we finally find, for all values of the classical field $\phi_c$, 
\be
\frac{dU_{eff}}{d\phi_c}=-\frac{k}{V^{(4)}}=0~,
\ee
such that the effective potential corresponding to the bare potential (\ref{cosine}) is a constant in the limit $N\to\infty$.

\end{document}